\documentstyle[12pt]{amsart}

\catcode`\@=11

\long\def\@savemarbox#1#2{\global\setbox#1\vtop{\hsize\marginparwidth 
  \@parboxrestore\tiny\raggedright #2}}
\newcommand\lref[1]{\ref{#1}%
\@ifundefined{r@DisplaY #1}{}{ (#1)}}

\newcommand\fakelabel[2]{\@bsphack\if@filesw {\let\thepage\relax
   \newcommand\protect{\noexpand\noexpand\noexpand}%
\xdef\@gtempa{\write\@auxout{\string
      \newlabel{#1}{{#2}{\thepage}}}}}\@gtempa
   \if@nobreak \ifvmode\nobreak\fi\fi\fi\@esphack}

\catcode`\@=12

\def\Empty{}
\newcommand\oplabel[1]{
  \def\OpArg{#1} \ifx \OpArg\Empty {} \else
        \label{#1}
  \fi}
\newtheorem{theoremSt}{Theorem}[section]

\newtheorem{exampleSt}[theoremSt]{Example}
\newtheorem{exerciseSt}[theoremSt]{Exercise}

\newcommand\MakeStEnv[1]{
  \newenvironment{#1}[1]{
  \begin{#1St} \oplabel{##1}%
  \global\def\CrntSt{\thetheoremSt}%
}{ 
  \end{#1St} }
  \newenvironment{#1+}[1]{
  \begin{#1St} \label{##1}%
  \label{DisplaY ##1}%
  \global\def\CrntSt{\thetheoremSt}%
  \def\Labl{##1}\ifx\Labl\Empty{} \else {\em (\Labl)\,}\fi%
}{ 
  \end{#1St} }
}
\MakeStEnv{theorem}
\MakeStEnv{corollary}
\MakeStEnv{proposition}
\MakeStEnv{lemma}
\MakeStEnv{definition}
\MakeStEnv{conjecture}

\newlength{\saveu}

\newcommand{\startproof}[1]{%
\medbreak\mbox{}\noindent{\it Proof of #1:}%
}
\newcommand{\finishproof}[1]{ 
  \def\FPArg{#1}
  \ifx\FPArg\Empty
        \newcommand\FPArg{\CrntSt}  \fi
  \smallbreak\noindent\makebox[\textwidth]{\hfill\fbox{\FPArg}}
  \medbreak\noindent
}

\newcommand\FF{{\cal F}}

\newcommand\LL{{\cal L}}
\newcommand\MM{{\cal M}}

\newcommand\PP{{\cal P}}

\newcommand\PMF{{\PP\kern-2pt\MM\FF}}

\newcommand\PML{{\PP\kern-2pt\MM\LL}}

\newcommand\bbR{{\mathord{\text{I\kern-2pt R}}}}        
\newcommand\bbH{{\mathord{\text{I\kern-2pt H}}}}        

\newcommand\bigrightarrow[1]{\hbox to #1{\rightarrowfill}}
\newcommand\bigleftarrow[1]{\hbox to #1{\leftarrowfill}}

\newcommand\semidir{\mathrel{\hbox{\vrule depth-.03ex height1.1ex\kern-0.15em$\times$}}}

\numberwithin{equation}{section}

\begin{document}

\title[Semiclassical Approximation for Chern-Simons Theory]
{Semiclassical Approximation for Chern-Simons Theory and 3-Hyperbolic
Invariants}

\author{A.A. Bytsenko}
\address{Departamento de Fisica, Universidade Estadual de Londrina,
Caixa Postal 6001, Londrina-Parana, Brazil; on leave from
Sankt-Petersburg State Technical University\,\, {\em E-mail address:} 
{\rm abyts@fisica.uel.br}}
\author{L. Vanzo}
\address{Dipartimento di Fisica, Universit\'{a} di Trento, and Istituto
Nazionale di Fisica Nucleare, Gruppo Collegato di Trento, Italy
\,\, {\em E-mail address:} {\rm vanzo@science.unitn.it}}
\author{S. Zerbini}
\address{Dipartimento di Fisica, Universit\'{a} di Trento, and Istituto
Nazionale di Fisica Nucleare, Gruppo Collegato di Trento, Italy
\,\, {\em E-mail address:} {\rm zerbini@science.unitn.it}}

\date{May, 1999}

\thanks{We thank Prof. F.L. Williams for useful discussion.
First author partially supported by a CNPq grant (Brazil), RFFI 
grant (Russia) No 98-02-18380-a, and by GRACENAS grant (Russia) 
No 6-18-1997.}

\maketitle

\begin{abstract}

The invariant integration method for Chern-Simons theory defined on the 
compact hyperbolic manifold $\Gamma\backslash {\Bbb H}^3$ is verified in the 
semiclassical approximation. The semiclassical limit for the partition 
function is presented. We discuss briefly $L^2-$ analytic torsion
and the eta invariant of Atiyah-Patodi-Singer for compact hyperbolic
3-manifolds.    

\end{abstract}

\vspace{1cm}

\section{Introduction}

It is known that topological invariants associated with $3$-manifolds can be
constructed within the framework of Chern-Simons gauge theory 
\cite{witt89-121-351}. These values have been specified in terms of the axioms 
of topological quantum field theory in \cite{moor89-123-77}, whereas 
the equivalent derivation of invariants has also been presented 
 combinatorially in
\cite{resh90-127-1,resh91-103-547}, where modular Hopf algebras related to 
quantum groups have been used. The Witten's (topological) invariants have been 
explicitly calculated for a number of 3-manifolds
and gauge groups
\cite{dijk90-129-393,kirb91-105-473,free91-141-79,jeff92-147-563,rama93-8-2285,
roza93u-99,roza96-175-275}. The semiclassical approximation for the 
Chern-Simons partition function may be expressed by
the asymptotics for  $k\rightarrow\infty$ of Witten's invariant of a 
$3$-manifold $M$ and a gauge group $G$. Typically this expression is a 
partition function of quadratic functional. This asymptotics leads to a series of
$C^{\infty}-$ invariants associated with triplets $\{M;F;\xi\}$ with $M$ a smooth
homology 
$3-$ sphere, $F$ a homology class of framings of $M$, and $\xi$ an acyclic 
conjugacy class of ortogonal representations of the fundamental group
 $\pi_1(M)$ 
\cite{axel94-39-173}. In addition the cohomology $H(M;Ad\,\xi)$ of $M$ 
with respect to the local system related to $Ad\,\xi$ vanishes. 

This note is an extension of the two previous  papers 
\cite{byts97-505-641,byts98-13}. Here our aim is to use again the invariant 
integration method 
\cite{schw79-64-233,adam98-417} in its simplest form in order to evaluate
 the semiclassical approximation in the 
Chern-Simons theory. We do this analysing the partition function related 
to compact hyperbolic 3-manifolds $\Gamma\backslash {\Bbb H}^3$, where
${\Bbb H}^3$ is the real hyperbolic space and $\Gamma$ is a co-compact 
discrete group of isometries (for details see Ref. \cite{byts96-266}).

We conclude this section introducing the Witten's invariant defined by
the partition function associated with  a Chern-Simons gauge
theory

$$
{\frak W}(k)=\int {\cal D}Ae^{ikCS(A)}\mbox{,}\hspace{1.0cm} k\in {\Bbb Z}
\mbox{.}
\eqno{(1.1)}
$$
The formal integration in (1.1) is one over the gauge fields $A$ in a trivial 
bundle, i.e. 1-forms on the 3-dimensional manifold $X_\Gamma$ with values 
in Lie algebra ${\rm g}$ of a gauge group ${\frak G}$.
The Chern-Simons functional $CS(A)$ can be considered as a function on 
a space of connections on a trivial principal bundle over a compact oriented
3-manifold $X_{\Gamma}$ given by

$$
CS(A)=\frac{1}{4\pi}\int_{X_\Gamma}\mbox{Tr}\left(A\wedge dA+\frac{2}{3}A
\wedge A\wedge A\right)\mbox{.}
\eqno{(1.2)}
$$
Let $X$ be a locally symmetric Riemannian manifold with negative sectional 
curvature. Its universal covering ${\widetilde   X}\rightarrow X$ is a 
Riemannian symmetric space of rank one.
The group of orientation preserving isometries ${\widetilde G}$ of 
${\widetilde X}$ is a
connected semisimple Lie group of real rank one and 
${\widetilde   X}={\widetilde   G}/{\widetilde   K}$,
where ${\widetilde   K}$ is a maximal compact subgroup of ${\widetilde   G}$. 
The fundamental group of $X$ acts by covering transformations on 
${\widetilde   X}$ and gives
rise to a discrete, co-compact subgroup $\Gamma \subset {\widetilde   G}$ 
such that $X=\Gamma\backslash {\widetilde   G}/{\widetilde   K}$. 
Let $G$ be a linear connected finite 
covering of ${\widetilde   G}$, the embedding 
$\Gamma\hookrightarrow {\widetilde   G}$ lifts
to an embedding $\Gamma\hookrightarrow G$. Let $K\subset G$ be a maximal
compact subgroup of $G$, then $X_{\Gamma}=\Gamma\backslash G/K$ is a compact
manifold.
Let ${\rm g}={\frak k}\oplus {\frak p}$ be a
Cartan decomposition of the Lie algebra ${\rm g}$ of $G$. Let
${\frak a}\subset{\frak p}$ be a 
one-dimensional subspace and $J=K\bigcap G_{{\frak a}}$ be the 
centralizer of ${\frak a}$ in $K$. Fixing a positive root system of 
$({\rm g},{\frak a})$ we have the Iwasawa decomposition 
${\rm g}={\frak k}\oplus {\frak a}\oplus {\frak n}$. For
$G=SO(n,1)$ \,$(n\in {\Bbb Z}_{+})$, $K=SO(n)$, and $J=SO(n-1)$. 
The corresponding symmetric space of non-compact type is the real hyperbolic
space ${\Bbb H}^n$ of sectional curvature $-1$. Its compact dual space is
the unit $n-$ sphere.

Since $CS(A)$ does not contain any metric on $X_\Gamma$, the quantity
${\frak W}(k)$ is expected to be metric independent, namely to be a 
(well-defined) topological invariant of $X_\Gamma$. Indeed, this fact has 
been proved in Refs. \cite{resh90-127-1,resh91-103-547}. 
In the limit $k\rightarrow\infty$, the asymptotics of the Witten's invariant
(semiclassical approximation of Eq. (1.1)), involves only a partition 
functions of quadratic functionals \cite{witt89-121-351}, namely

$$
\sum_{[A_f]}\exp\left(ikCS(A_f)\right)\int {\cal D}\omega 
\exp\left(\frac{ik}{4\pi}\int_{{X_\Gamma}}
\mbox{Tr}(\omega\wedge d_{A_f}\omega)\right)
\mbox{.}
\eqno{(1.3)}
$$
In above equation the sum is taken over representatives $A_f$ for each point
$[A_f]$ in the moduli-space of flat gauge fields on $X_\Gamma$. In addition the
$\omega$ are Lie-algebra-valued 1-forms and $d_{A_f}$ is the covariant 
derivative determined by $A_f$, 

$$
d_{A_f}\omega=d\omega+[A_f,\omega]\mbox{.}
\eqno{(1.4)}
$$

\section{Quadratic functional with elliptic resolvent}

Let $M$ be a compact oriented Riemannian manifold without boundary, and $n=2m+1
=\mbox{dim}M$ is the dimension of the manifold.
Let $\chi:\pi_1(M)\longmapsto O(V,\langle\cdot\,,\cdot\rangle_V)$ be a
representation of $\pi_1(M)$ on real vectorspace $V$.
The mapping $\chi$ determines (on a basis of standard construction in
differential geometry) a real flat vectorbundle $\xi$ over $M$ and a flat
connection map $\nabla_p$ on the space $\Omega^p(M,\xi)$ of differential 
$p-$ forms on $M$ with values in $\xi$. One can say that $\chi$ determines 
the space of smooth sections in the vectorbundle 
$\Lambda^p(TM)^{*}\otimes \xi$. One can
construct from the metric on $M$ and Hermitian structure in $\xi$ a Hermitian
structure in $\Lambda(TM)^{*}\otimes\xi$ and the inner products $\langle
\cdot\,,\cdot\rangle_m$ in the space $\Omega^m(M,\xi)$. Thus

$$
S_{\cal O}=\langle\omega,{\cal O}\omega\rangle_m
\mbox{,}
\hspace{1.0cm}
{\cal O}=*\nabla_m
\mbox{,}
\eqno{(2.1)}
$$
where $(*)$ is the Hodge-star map. The map $\cal O$ is formally
self-adjoint with the property ${\cal O}^2=\nabla_m^{*}\nabla_m$.
Suppose that the quadratic functional (2.1) is
defined on the space ${\cal G}={\cal G}(M,\xi)$ of smooth sections in a
real Hermitian vectorbundle $\xi$ over $M$. 
There exists a canonical topological elliptic resolvent $R(S_{\cal O})$,
related to the functional (2.1), namely

$$
0\stackrel{0}{\longrightarrow}\Omega^0(M,\xi)\stackrel{\nabla_0}
{\longrightarrow}...
\stackrel{\nabla_{m-2}}{\longrightarrow}\Omega^{m-1}(M,\xi)
\stackrel{\nabla_{m-1}}
{\longrightarrow}{\rm ker}(S_{\cal O})\stackrel{0}{\longrightarrow}0
\mbox{.}
\eqno{(2.2)}
$$
Therefore, for the resolvent $R(S_{\cal O})$, we have 
${\cal G}_p=\Omega^{m-p}(M,\xi)$ and
$H^p(R(S_{\cal O}))=H^{m-p}(\nabla)$, where 
$H^p(\nabla)={\rm ker}(\nabla_p)/{\Im}(\nabla_{p-1})$ 
are the cohomology space. 
Note that $S_{\cal O}\geq 0$ and therefore
${\rm ker}(S_{\cal O})\equiv {\rm ker}({\cal O})={\rm ker}(\nabla_m)$.

Let us choose an inner product $\langle\cdot\,,\cdot\rangle_{H^p}$ in each
space $H^p(R(S_{\cal O}))$. The partition function of $S_{\cal O}$ with the
resolvent (2.2) can be written in the form (see Refs. \cite{adam95u-95,
adam98-417})

$$
{\frak W}(k)\equiv{\frak W}(k;R(S_{\cal O}),\langle\cdot\,,
\cdot\rangle_H,\langle\cdot\,,\cdot\rangle)
=\left(\frac{\pi}{k}\right)^{\zeta(0,|{\cal O}|)/2}
e^{-\frac{i\pi}{4}\eta(0,{\cal O})}
$$
$$
\times\tau(M,\chi,\langle\cdot\,,\cdot\rangle_H)^{1/2}
\mbox{,}
\eqno{(2.3)}
$$
where $|{\cal O}|=\sqrt{{\cal O}^2}$ is
defined via spectral theory. 
This is the basic formula one has to evaluate. 
With regard to the quantity $ \tau(M,\chi,\langle\cdot\,,\cdot\rangle_H) $,
it is related the Ray-Singer torsion. In fact, if 
$H^0(\nabla)\neq 0$ and $H^p(\nabla)= 0$ for
$p=1,...,m$, then the product

$$
\tau(M,\chi,\langle\cdot\,,\cdot\rangle_H)= T_{an}^{(2)}(M)
\cdot {\rm Vol}(M)
^{-{\rm dim} H^0(\nabla)}
\mbox{,}
\eqno{(2.4)}
$$
is metric independent \cite{ray71-7}, i.e. the metric dependence
of the Ray-Singer torsion $ T_{an}^{(2)}(M)$ factors out as
$V(M)^{-\mbox{dim}H^0(\nabla)}$.

As far as the zeta-function $\zeta(0,|{\cal O}| $  is concerned, we recall 
that there exists $\varepsilon,\delta >0$
such that for $0<t<\delta$ the heat kernel expansion for self-adjoint
Laplace operators ${\frak L}_p$ is given by

$$
\mbox{Tr}\left(e^{-t{\frak L}_p}\right)=\sum_{0\leq \ell\leq \ell_0} a_\ell
({\frak L}_p)t^{-l}+ O(t^\varepsilon)
\mbox{.}
\eqno{(2.5)}
$$
Starting with the formula \cite{adam95u-95}
$$
\zeta(0,{\frak L}_p)=a_0({\frak L}_p)-{\rm dim}({\rm ker}({\frak L}_p))
=a_0({\frak L}_p)
-{\rm dim}H^p(R(S))
\mbox{,}
\eqno{(2.6)}
$$
one can shown that the zeta function $\zeta(s,|{\cal O}|)$ 
 is well-defined and analytic for $\Re(s)>0$ and
can be continued to a meromorphic function on ${\Bbb C}$, regular at $s=0$ and

$$
\zeta(0,|{\cal O}|)=\sum_{p=0}(-1)^p(a_0({\frak L}_p)-{\rm dim} H^p(R(S)))
\mbox{.}
\eqno{(2.7)}
$$
Furthermore, the zeta function $\zeta(0,|{\cal O}|)$ appearing in the 
partition  function 
(2.3) can be expressed in terms of the dimensions of the cohomology spaces 
of ${\cal O}$. Indeed, if the dimension of $M$ is odd $(n=2m+1)$ then for all 
$p$\, $a_0({\frak L}_p)=0$, because we are dealing with manifold without 
boundary. Since 
$H^p(R(S_{\cal O}))=H^{m-p}(\nabla)$ (the Poincar{\`e} duality) 
for the resolvent (2.2), 
it follows that 

$$
\zeta(0||{\cal O}|)=-\sum_{p=0}^m (-1)^p{\rm dim} H^p(R(S))=(-1)^{m+1}
\sum_{p=0}^m (-1)^p{\rm dim} H^p({\nabla})
\mbox{.}
\eqno{(2.8)}
$$

Finally, the dependence of the eta invariant $\eta(0|{\cal O})$ of 
Atiyah-Patodi-Singer 
on the connection map ${\cal O}$
can be expressed with the help of the formula for the index of the twisted
signature operator for a certain vectorbundle over $M\otimes[0,1]$
(see \cite{atiy75-77,atiy75-78,atiy76-79}). Furthermore It can be shown
 \cite{adam95u-95}
that

$$
\eta(s|B)=2\eta(s|{\cal O})
\mbox{,}
\eqno{(2.9)}
$$
where the $B$ are elliptic self-adjoint maps on $\Omega(M,\xi)$ 
defined on $p$-forms by

$$
B_p=(-i)^{\lambda(p)}\left(*\nabla+(-1)^{p+1}\nabla*\right)
\mbox{,}
\eqno{(2.10)}
$$
In this formula $\lambda(p)=(p+1)(p+2)+m+1$ and for the Hodge star-map we have
used $*\alpha\wedge\beta=\langle\alpha,\beta\rangle_{vol}$. From the Hodge
theory we have
$$
{\rm dim}\mbox{ker}B=\sum_{p=0}^m{\rm dim}H^p(\nabla)
\mbox{.}
\eqno{(2.11)}
$$

\section{The case of real compact hyperbolic manifolds}

In this section, we shall consider the specific case of a  compact 
  hyperbolic 3-manifolds of the form 
$M=X_{\Gamma}=\Gamma\backslash {\Bbb H}^3$. 
If the flat bundle, $\xi$ is acyclic, then for $L^2-$ torsion one gets
\cite{frie86-84}: 
$[T_{an}^{(2)}(X_\Gamma)]^2={\frak R}_{\chi}(0)$, where 
${\frak R}_{\chi}(s)$ is
the Ruelle function. The function ${\frak R}_{\chi}(s)$ is an alternating 
product of more complicate 
factors, each of which is a Selberg zeta function $Z_p(s;\chi)$.
The relation of Ruelle and Selberg functions is: 

$$
{\frak R}_{\chi}(s)=\prod_{p=0}^{{\rm dim}M-1}Z_p(p+s;\chi)^{(-1)^j}
\mbox{.}
\eqno{(3.1)}
$$
The function ${\frak R}_{\chi}(s)$ extends meromorphically to the entire 
complex plane $\Bbb C$ \cite{deit89-59}.
The Ruelle function associated with closed oriented 
hyperbolic 3-manifold $X_\Gamma=\Gamma \backslash {\Bbb H}^3$ has the form
${\frak R}_{\chi}(s)=Z_0(s;\chi)Z_2(2+s;\chi)/Z_1(1+s;\chi)$.
The analytic torsion for manifold $X_\Gamma$ has been calculated (in the
presence of non-vanishing Betti numbers $b_i\equiv b_i(X_\Gamma)=
{\rm rank}_{\Bbb Z}H_i(X_\Gamma;{\Bbb Z})$) in Refs. 
\cite{byts97-505-641,byts98-13}.

Now we consider the evaluation of eta invariant contribution. 
With regard to this point, a remarkable formula relating 
$\eta(s,{\cal O})$ to the closed geodesics on $X_\Gamma$ 
has been obtained by Millson \cite{mill78-108}. More explicitly, Millson 
has proved the following result for a Selberg type (Shintani) zeta function
${\widetilde Z}(s,{\cal O})$.

Let us define a zeta function by the following series, which is absolutely 
convergent for $\Re(s)>0$,

$$
{\rm log}{\widetilde Z}(s,{\cal O})\stackrel{def}{=}\sum_{[\gamma]\neq 1}
\frac{{\rm Tr}\tau^{+}_{\gamma}-{\rm Tr}\tau^{-}_{\gamma}}
{|{\rm det}(I-P_h(\gamma))|^{1/2}}\frac{e^{-s\ell(\gamma)}}{m(\gamma)}
\mbox{,}
\eqno{(3.2)}
$$
where $[\gamma]$ runs over the nontrivial conjugacy classes in 
$\Gamma=\pi_1(X_\Gamma)$, $\ell(\gamma)$ is the length of the closed geodesic 
$c_{\gamma}$ (with multiplicity $m(\gamma)$) in the free homotopy class
corresponding to $[\gamma]$, $P_h(\gamma)$ is the restriction of the linear
Poincar\'{e} map $P(\gamma)=d\Phi_1$ at 
$(c_{\gamma},\dot{c}_{\gamma})\in TX_\Gamma$
to the directions normal to the geodesic flow $\Phi_t$ and 
$\tau^{\pm}_{\gamma}$ is the parallel translation around $c_{\gamma}$ on
$\Lambda^{\pm}_{\gamma}=\pm i$ eigenspace of 
$\sigma_B(\dot{c}_{\gamma})$ ($\sigma_B$ denoting the principal symbol of 
${\cal O}$). Then ${\widetilde Z}(z,{\cal O})$ admits a meromorphic 
continuation to the entire complex plane, which in particular is 
holomorphic at $s=0$ and

$$
{\rm log}{\widetilde Z}(0,{\cal O})=\pi i\eta(0,{\cal O})
\mbox{.}
\eqno{(3.3)}
$$
Furthermore, it is possible to show that  ${\widetilde Z}(s,{\cal O})$
satisfies the functional equation

$$
{\widetilde Z}(s,{\cal O}){\widetilde Z}(-s,{\cal O})=
e^{2\pi i\eta(0,{\cal O})}
\mbox{.}
\eqno{(3.4)}
$$

Now we have all the ingredients for the evaluation of the partition 
function (2.3) in terms of $L^2-$ analytic
torsion and a Selberg type function. The final result is

$$
{\frak W}(k)=\left(\frac{\pi}{k}\right)^{\zeta(0,|{\cal O}|)/2}
{\widetilde Z}(0,{\cal O})^{-1/4}
\left[T_{an}^{(2)}(X_\Gamma)\right]^{1/2}
\left[{\rm Vol}(\Gamma\backslash G)\right]^{-{\rm dim}H^0(\nabla)/2}
\mbox{,}
\eqno{(3.5)}
$$
where $\zeta(0,{\cal O})$ and ${\widetilde Z}(0,{\cal O})$ are given by 
Eqs. (2.8) and (3.2) respectively.

\section{Concluding remarks}

For a real compact hyperbolic 3-manifold, the  formula (3.5) gives the 
value of the
asymptotics of the Chern-Simons-Witten invariant. This is the main 
result of our paper.
The invariant (3.5) involves the $L^2-$ analytic torsion, which can be 
expressed by means of Selberg zeta functions and a Shintani zeta function 
${\widetilde Z}(0,{\cal O})$ associated with the
eta invariant of Atiyah-Patodi-Singer \cite{atiy75-77}. Finally we note that
the explicit result (3.5) can be very important for investigating  
the relation between
quantum invariants for an oriented 3-manifold, defined with the help of a 
representation theory of quantum groups \cite{resh90-127-1,resh91-103-547},
and Witten's invariant \cite{witt89-121-351}, which is, instead, related to 
the path integral approach.

\end{document}